\documentclass[a4paper,fleqn]{cas-sc}

\usepackage[authoryear]{natbib}
\usepackage{booktabs}
\usepackage{multirow}
\usepackage{subcaption} 
\usepackage{float}
\usepackage{placeins}


\def\tsc#1{\csdef{#1}{\textsc{\lowercase{#1}}\xspace}}
\tsc{WGM}
\tsc{QE}

\begin{document}
\let\WriteBookmarks\relax
\def\floatpagepagefraction{1}
\def\textpagefraction{.001}

\shorttitle{}    

\shortauthors{M. Jeong et al.}  

\title [mode = title]{AgentSUMO: An Agentic Framework for Interactive Simulation Scenario Generation in SUMO via Large Language Models}  




\author[1]{Minwoo Jeong}[orcid=0009-0005-1025-0294]

\author[1]{Jeeyun Chang}[orcid=0009-0002-1706-5946]

\author[1,2]{Yoonjin Yoon}[orcid=0000-0002-3550-4431]
\cormark[1]
\ead{yoonjin@kaist.ac.kr}

\affiliation[1]{organization={Graduate School of Data Science, Korea Advanced Institute of Science and Technology},
            city={Daejeon}, 
            country={Republic of Korea}}
            
\affiliation[2]{organization={Department of Civil and Environmental Engineering, Korea Advanced Institute of Science and Technology},
            city={Daejeon}, 
            country={Republic of Korea}}

\cortext[cor1]{Corresponding author}


\begin{abstract}
The growing complexity of urban mobility systems has made traffic simulation indispensable for evidence-based transportation planning and policy evaluation. However, despite the analytical capabilities of platforms such as the Simulation of Urban MObility (SUMO), their application remains largely confined to domain experts. Developing realistic simulation scenarios requires expertise in network construction, origin–destination modeling, and parameter configuration for policy experimentation, creating substantial barriers for non-expert users such as policymakers, urban planners, and city officials. Moreover, the requests expressed by these users are often incomplete and abstract—typically articulated as high-level objectives, which are not well aligned with the imperative, sequential workflows employed in existing language-model-based simulation frameworks. To overcome these challenges, this study proposes AgentSUMO, an agentic framework for interactive simulation scenario generation via large language models. AgentSUMO departs from imperative, command-driven execution by introducing an adaptive reasoning layer that interprets user intents, assesses task complexity, infers missing parameters, and formulates executable simulation plans. The framework is structured around two complementary components: the Interactive Planning Protocol, 
which governs reasoning and user interaction, and the Model Context Protocol, which manages standardized communication and orchestration among simulation tools. Through this design, AgentSUMO converts abstract policy objectives into executable simulation scenarios. Experiments on urban networks in Seoul and Manhattan demonstrate that the agentic workflow achieves substantial improvements in traffic flow metrics 
while maintaining accessibility for non-expert users, successfully bridging the gap between high-level policy goals and executable simulation workflows.
 
\end{abstract}




\maketitle

\section{Introduction}\label{sec:intro}
Rapid urbanization and shifting mobility patterns are intensifying pressure on urban transportation networks worldwide \citep{MA2025105311}. By 2050, more than 68\% of the global population is projected to live in cities, and traffic congestion already costs the U.S. economy nearly \$180 billion annually, with similar trends observed globally~\citep{UN_DESA_2018_Urbanization, CNBC_2019_TrafficJams}. To mitigate these growing challenges, cities are increasingly deploying Intelligent Transportation Systems (ITS) that leverage advanced sensing, communication, and control technologies to enhance network efficiency and reliability~\citep{s21062143, KUANG2025105325}. These ITS deployments have become central to broader smart-city initiatives, integrating with urban data platforms and IoT infrastructures to form responsive, interconnected mobility ecosystems—thereby creating a pressing need for advanced modeling and simulation tools capable of evaluating their effectiveness under realistic conditions.

As cities adopt advanced ITS and smart-city initiatives, traffic simulation has become indispensable for evaluating algorithms and policy interventions before real-world deployment~\citep{s21062143}. 
By replicating transportation dynamics through mathematical and computational models, simulation supports planning, design, and operational decisions across diverse urban contexts. 
It enables engineers and researchers to analyze adaptive signal control, multimodal coordination, and network performance under varying conditions, while also providing policymakers and planners with a safe, cost-effective means to assess infrastructure strategies, congestion-mitigation measures, and emerging mobility policies \citep{Dorokhin_2020}. 
Such virtual experimentation reduces both financial risk and implementation uncertainty, making simulation a critical foundation for evidence-based urban mobility management.

Despite the growing importance of traffic simulation, substantial accessibility barriers persist for non-expert users such as urban planners and policymakers \citep{PELL20171477}. 
Traffic simulators differ widely in modeling granularity—from macroscopic flow-based representations to mesoscopic approximations and detailed microscopic models. 
While highly granular tools such as the Simulation of Urban MObility (SUMO) offer powerful capabilities for detailed analysis~\citep{behrisch2011sumo}, their operational complexity results in a steep learning curve. Building realistic simulation scenarios is often a technical and time-intensive process that requires network extraction from OpenStreetMap (OSM), preparation of origin–destination (OD) matrices, configuration of multiple XML files for routes and control schemes, and extensive post-processing of simulation outputs \citep{kotusevski2009review}. 
These demanding steps hinder rapid prototyping and iterative analysis, creating a persistent gap between the analytical potential of simulation platforms and the practical needs of decision makers who require efficient, interpretable, and user-friendly tools.

\begin{figure}[pos=t]
  \centering
  \includegraphics[width=\textwidth,height=0.6\textheight,keepaspectratio]{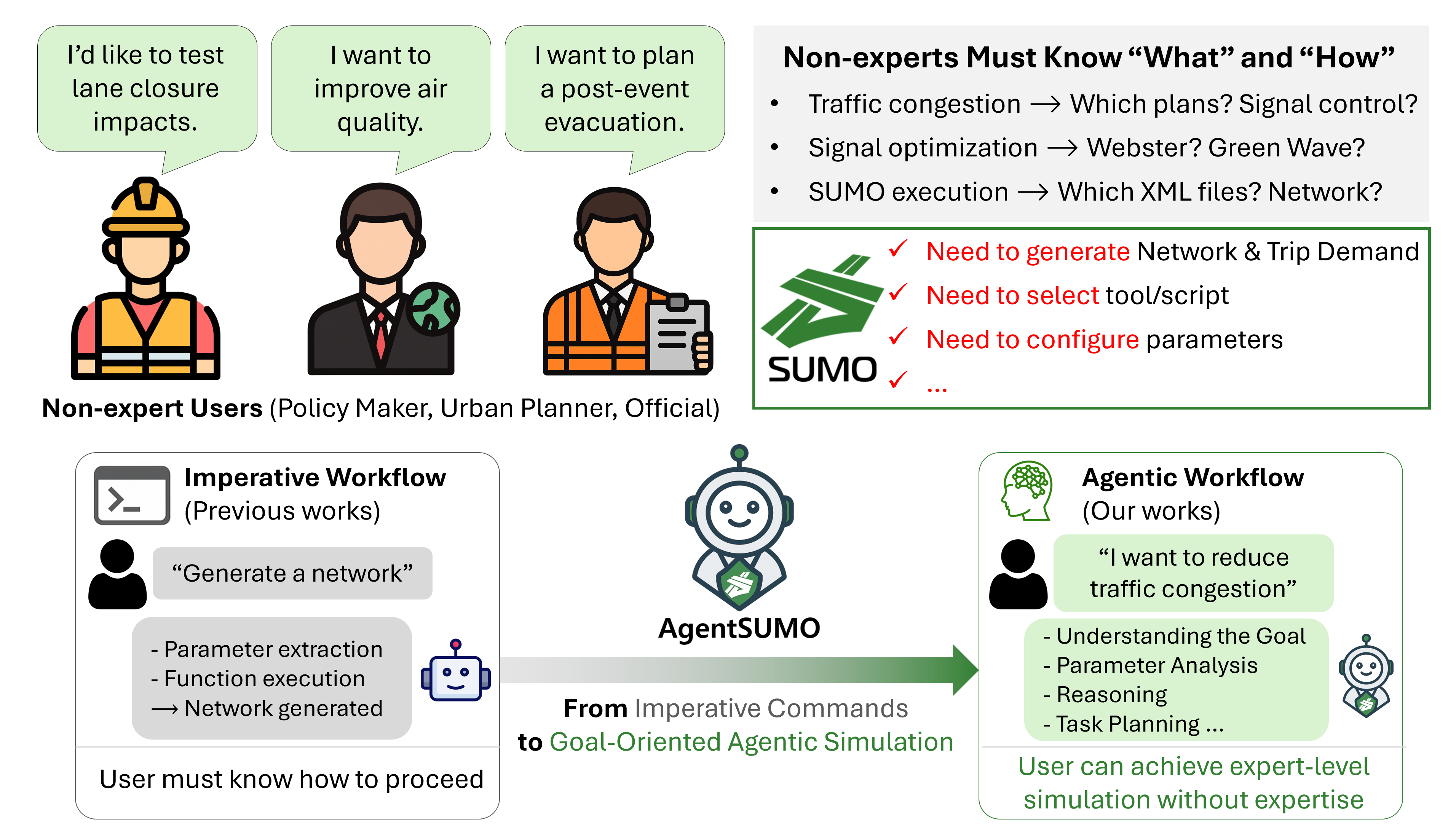}
  \caption{Illustration of the Motivation for AgentSUMO. The framework enables non-expert users to perform expert-level traffic simulations by converting high-level goals (e.g., “reduce congestion”) into executable SUMO workflows without manual scripting.}
  \label{fig:motivation}
\end{figure}

Recent advances in large language models (LLMs) have opened new opportunities to overcome technical barriers that limit the accessibility of traffic simulation.
By combining natural-language understanding, multi-step reasoning, and autonomous task planning, LLMs can interpret complex user instructions and automate processes that previously required extensive manual scripting \citep{minaee2024large, huang2024understanding}. These capabilities are further strengthened by frameworks such as LangChain, which integrate external tools, manage memory, and orchestrate workflows \citep{topsakal2023creating}. Building upon these technological foundations, recent studies have begun integrating LLMs with traffic simulation to enable natural-language interaction and streamline scenario setup. Approaches such as \textit{TrafficGPT}, \textit{LLMLight}, \textit{ChatSUMO}, and \textit{SUMO-MCP} have shown promise in enabling conversational control of simulation tasks including traffic data analysis, signal optimization, and network generation within SUMO~\citep{ZHANG202495, lai2025llmlight, li2024chatsumo, ye2025sumo}. 

While these methods successfully lower the entry barrier by translating natural-language queries into simulation commands, most remain confined to an \textbf{imperative workflow}. 
In this paradigm, users must explicitly instruct the system to execute discrete commands—such as “generate a network,” “configure signals,” or “run simulation”—and still possess sufficient technical expertise to determine which specific procedures are required to achieve higher-level objectives like congestion reduction or emission mitigation. 
As a result, current LLM-integrated SUMO frameworks assist users in expressing \textit{how} to perform tasks but not in reasoning \textit{what} should be done to accomplish broader policy goals.

To bridge this gap between simulation capabilities and decision-making needs, this paper proposes \textbf{AgentSUMO}, an agentic framework for interactive traffic simulation in SUMO via LLMs.

Figure~\ref{fig:motivation} illustrates this paradigm shift. 
Previous approaches follow an imperative workflow, where users must explicitly instruct each simulation step, such as generating networks, configuring signals, or defining OD matrices. As a result, non-expert users—such as urban planners and policymakers—are required to understand both the \textit{what} and \textit{how} of simulation processes, even though their queries are often incomplete or abstract (e.g., “reduce congestion” or “improve air quality”). In contrast, AgentSUMO introduces a goal-oriented agentic workflow in which the LLM autonomously interprets such high-level intents, infers missing parameters, and decomposes the user’s request into executable SUMO operations. This transition from command-driven to reasoning-driven interaction enables users to achieve expert-level simulation and policy evaluation without detailed technical knowledge.

AgentSUMO unifies LLM reasoning with traffic simulation to turn high-level policy intents into executable SUMO studies. 
Instead of asking users to issue stepwise commands, the agent conducts a clarify-before-execute dialogue, derives the missing parameters, and plans the simulation tasks end-to-end. 
It is organized around two complementary layers: the Interactive Planning Protocol, which manages adaptive reasoning and user interaction, and the Model Context Protocol (MCP), which coordinates standardized tool execution and data exchange across simulation modules. 
Through this integration, AgentSUMO orchestrates network and route generation, performs controlled comparisons across policies, and delivers interpretable summaries and visualizations. 
By coupling adaptive reasoning with protocol-based orchestration, the framework enables non-experts to design, execute, and evaluate realistic simulations directly from abstract goals.

The main contributions of this study are as follows:
\begin{itemize}
    \item We develop AgentSUMO, an agentic traffic simulation framework that enables goal-oriented, end-to-end orchestration of SUMO workflows through natural-language interaction. 
    \item We design an Interactive Planning Protocol for adaptive reasoning and MCP for standardized tool integration, allowing dynamic scenario generation, routing, and simulation analysis. 
    \item We demonstrate AgentSUMO’s effectiveness through multi-level experiments showing its ability to autonomously reason about simulation objectives and evaluate traffic-management policies under realistic urban conditions. 
    \item We conduct a systematic evaluation illustrating how the agentic workflow lowers the technical barrier for non-experts while maintaining analytical rigor comparable to expert-designed simulations.
\end{itemize}

The remainder of this paper is organized as follows. 
Section~\ref{sec:related} reviews the technical foundations of agentic AI and recent advances in LLMs relevant to transportation research and traffic simulators. 
Section~\ref{sec:method} introduces the architecture and agentic workflow of AgentSUMO, detailing how the Interactive Planning Protocol and MCP jointly enable adaptive reasoning, standardized tool orchestration, and end-to-end simulation execution. 
Section~\ref{sec:results} presents experimental evaluations conducted on urban networks in Seoul and Manhattan, illustrating how the framework performs across real-world scenarios to support realistic, policy-relevant traffic analyses. 
Finally, Section~\ref{sec:conclusion} summarizes the key findings and discusses future directions.

\section{Related Works}\label{sec:related}

\subsection{Agentic AI and LLMs: Technical Foundations}
The emergence of Large Language Models (LLMs) has transformed their role from simple text generators into sophisticated reasoning systems. This shift has been accelerated by advances in prompt-level techniques such as Chain-of-Thought (CoT) \citep{wei2022chain}, which enables step-by-step problem decomposition; ReACT \citep{yao2023react}, which integrates reasoning with action execution; Self-Consistency \citep{10298561}, which aggregates multiple reasoning paths to improve accuracy; and Tree-of-Thoughts (ToT) \citep{NEURIPS2023_271db992}, which frames reasoning as a search process over thought trees. Complementary to these reasoning advances, mechanisms such as prompt caching have optimized LLM usage through reducing redundant computation \citep{gim2024prompt}. Collectively, these advanced prompting methods demonstrate that LLMs can operate as efficient reasoners capable of inference and planning, extending far beyond their original role as language generators.

Building on these reasoning capabilities, the introduction of LangChain provided a foundational framework for orchestrating LLM-based agents through tool integration, memory management, and workflow coordination \citep{topsakal2023creating}. This enabled researchers and practitioners to design LLMs as autonomous agents that interact with external environments, maintain context across interactions, and perform multi-step tasks. Subsequently, the field has seen a rapid proliferation of diverse agent frameworks. AutoGen \citep{wu2023autogenenablingnextgenllm} demonstrated the potential of autonomous, goal-driven agents with minimal human intervention; CrewAI introduced role-based multi-agent collaboration patterns; LangGraph modeled agent behavior as graphs with explicit states and transitions, supporting branching logic, iterative loops, and complex control flows \citep{duan2024explorationllmmultiagentapplication, HOSSEINI2025100399}. These developments have broadened the design space for agentic architectures in different domains.

Recently, attention has shifted toward standardization and readiness. Anthropic’s MCP exemplifies this trend by defining a unified protocol for connecting LLMs with external tools \citep{hou2025modelcontextprotocolmcp}. Taken together, these developments illustrate the evolution of agentic AI—from early prompt-based reasoning, through the experimental phase of framework innovation, to the emergence of stable, standardized systems capable of complex autonomous operation \citep{10849561}.

\subsection{Advancements in LLMs for Transportation}
Recent advancements in LLMs have extended their applications across diverse areas of transportation research. Within the ITS domain, LLMs are now used for various purposes such as traffic forecasting, signal control optimization, driver assistance and safety analysis, and broader urban traffic management \citep{10851302, long2025survey}.
TPLLM \citep{ren2024tpllmtrafficpredictionframework} leverages LLM-based reasoning to model temporal and spatial dependencies in large-scale traffic data, achieving accurate short- and long-term flow predictions through prompt-driven forecasting. TrafficGPT \citep{ZHANG202495} bridges LLM reasoning with existing traffic foundation models, enabling interactive analysis, summarization, and multi-scale interpretation of spatio-temporal traffic data. UrbanGPT \citep{10.1145/3637528.3671578} represents a city-scale spatio-temporal foundation model that directly learns and predicts urban phenomena—such as taxi flow and congestion—through instruction-tuned LLM architectures. Frameworks like LA-Light \citep{wang2024llmassistedlightleveraginglarge} and LLMLight \citep{lai2025llmlight} use LLM-driven reasoning to adapt signal timing based on contextual cues such as vehicle density, time of day, and external conditions, yielding transparent and data-responsive control policies. Vision-language and LLM-augmented systems such as TrafficSafetyGPT \citep{zheng2023trafficsafetygpttuningpretrainedlarge}, ChatScene \citep{Zhang_2024_CVPR}, AccidentGPT \citep{wu2024accidentgptlargemultimodalfoundation}, and DriveLLM-V \citep{HOU2025105368} apply natural-language reasoning to interpret or generate complex driving scenes, anticipate hazards, and support safer maneuver planning.
Collectively, these studies show that LLMs are increasingly embedded throughout the ITS ecosystem—from prediction and control to situational understanding and decision support.

\subsection{LLM Integration with Traffic Simulators}
Building on the natural-language interaction paradigm, recent research has focused on integrating LLMs directly with traffic simulation platforms. ChatSUMO connects LLM to SUMO enabling end-to-end natural language control over simulation workflows—from scenario creation to customization and analysis. Users can define locations, network sizes, and traffic intensities conversationally, and the system translates these descriptions into executable SUMO configurations. Its imperative, pipeline-based architecture—comprising input, generation, customization, and analysis modules—ensures predictable and reproducible outcomes. However, this structured design also constrains flexibility, making it difficult to explore alternative workflows or perform complex experimentation without predefined steps.

Conversely, SUMO-MCP adopts a protocol-based orchestration strategy. By leveraging Anthropic’s MCP, it allows dynamic sequencing of simulation tools in response to user prompts, supporting adaptive and exploratory simulation flows. Yet, SUMO-MCP emphasizes protocol compliance over conversational richness, therefore lacking the interactive dialogue loop that enables users iteratively refine scenarios and request explanations.

While ChatSUMO demonstrated accessibility through imperative pipelines and SUMO-MCP demonstrated flexibility through protocol-driven orchestration, each system remains limited on the other axis. To overcome these limitations, we propose AgentSUMO, a framework that combines conversational accessibility with orchestration adaptability. By supporting multi-turn reasoning, iterative scenario refinement, and policy experimentation through natural dialogue, our framework advances toward a truly agentic simulation environment—where LLMs not only execute SUMO workflows but also engage with users to co-develop, evaluate, and refine complex urban mobility scenarios. 

\begin{figure}[pos=t]
  \centering
  \includegraphics[width=\linewidth,height=0.8\textheight,keepaspectratio]{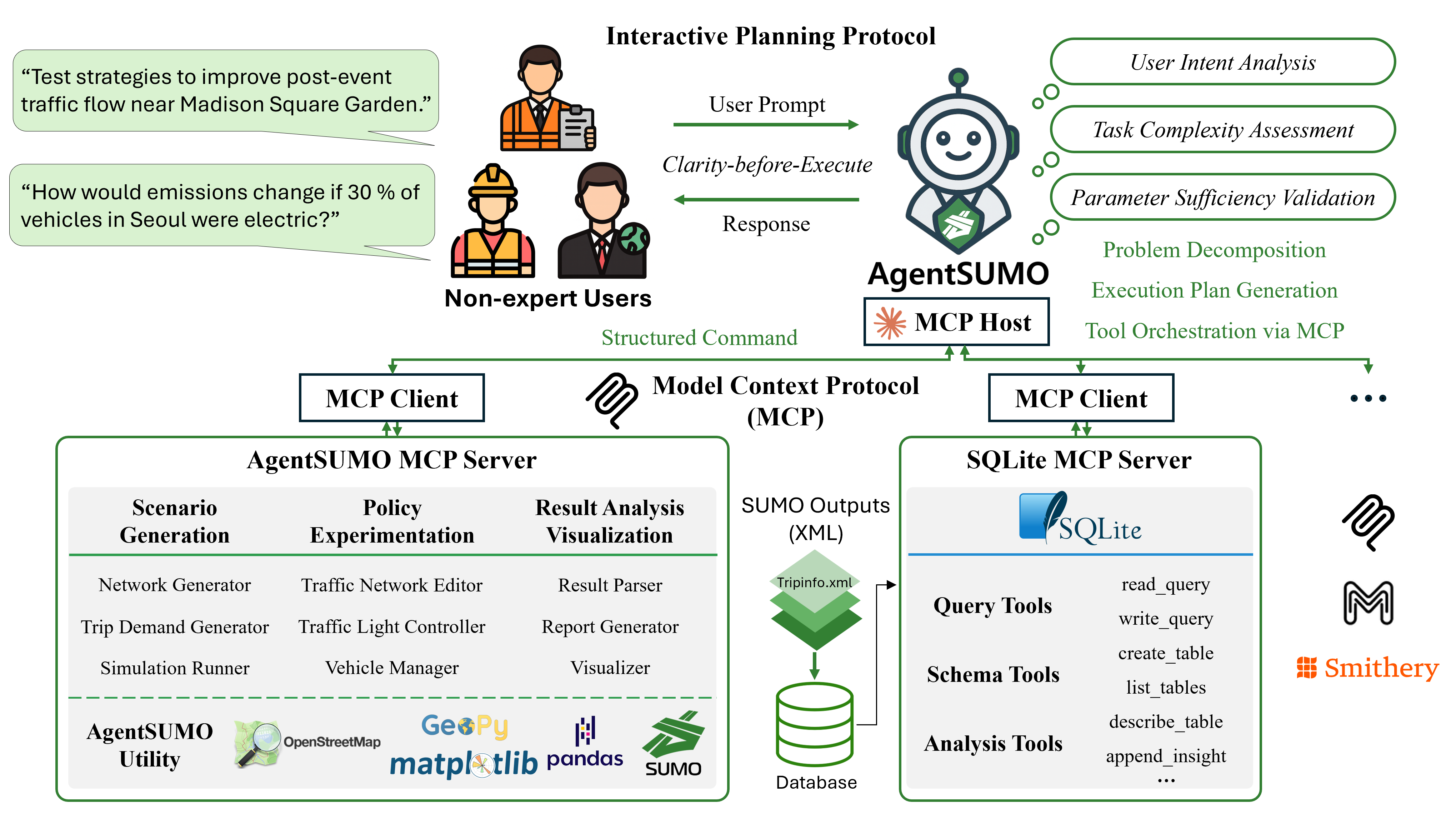}
  \caption{Overall framework of AgentSUMO.} 
  \label{fig:method_overview}
\end{figure}

\section{Methodology}\label{sec:method}
\subsection{Overview}
AgentSUMO is an agentic traffic simulation framework that bridges LLM reasoning with the SUMO, enabling non-expert users to design and execute traffic simulations through natural-language interaction. As shown in Figure \ref{fig:method_overview}, the framework integrates two complementary layers: the Interactive Planning Protocol that governs reasoning and user interaction, and the MCP that standardizes tool orchestration and data exchange across simulation components.
Through the Interactive Planning Protocol, the Planner Agent interprets high-level user intents, assesses task complexity, validates parameter sufficiency, and transforms abstract goals into executable simulation plans.
Once a scenario is confirmed, the agent decomposes the problem, generates an executable plan, and orchestrates domain-specific tools via the MCP.
Structured commands are transmitted to the AgentSUMO MCP Server, which coordinates tool invocations for scenario generation, policy experimentation, result analysis and visualization.
Simulation outputs are then converted into structured databases through an SQLite MCP Server, supporting efficient querying and cross-scenario comparison.
This design realizes an end-to-end agentic simulation workflow in which reasoning, execution, and analysis are seamlessly connected—from user prompt to interpretable insight.
Building on our previous work \citep{jeong2025speak}, the present study extends the framework by redesigning the model architecture through the MCP, introducing adaptive reasoning via the Interactive Planning Protocol, and broadening evaluation across diverse and realistic policy scenarios and urban networks.

\subsection{Planner Agent Design}
The Planner Agent serves as the orchestrator that converts natural-language intents into executable SUMO workflows. We assume that non-expert queries are often incomplete and abstract, making direct execution of traffic simulations challenging. The agent therefore collaborates with the user to co-construct simulation scenarios and ensures that each execution is safe and reproducible. Unlike imperative systems that require explicit procedural commands, our agent supports goal-oriented interaction, in which users specify desired outcomes and the agent autonomously plans and performs the necessary workflow. Conventional imperative pipelines can interpret natural language but still depend on explicit commands for each execution step and lack agentic planning capability. To bridge this gap, we design an Interactive Planning Protocol and embed it within the system prompt. This protocol enables the agent to interpret user intent, adapt its reasoning depth to task complexity, and confirm plans before execution, thereby facilitating accessible scenario design for non-expert users.

\subsubsection{Interactive Planning Protocol}\label{sec:interactive-protocol}

The Interactive Planning Protocol governs how the Planner Agent interprets user intent and transforms under-specified queries into executable SUMO scenarios. Non-expert users often express goals in abstract or incomplete form (e.g., ``reduce congestion near Gangnam Station''), which traditional pipelines cannot execute without extensive configuration. To overcome this limitation, the protocol embeds structured decision logic within the system prompt, enabling consistent interpretation, reasoning, and execution across all simulation requests. 
Prompt engineering is the process of designing natural-language instructions to guide LLMs without additional training or fine-tuning~\citep{sahoo2024systematic}. 
Recent studies have demonstrated that carefully structured prompts alone can induce multi-step reasoning, tool-use planning, and adaptive decision behavior in LLMs~\citep{wei2022chain, zhou2022least, yao2023react, huang2022language}. 
Building on these findings, our Interactive Planning Protocol leverages system-level prompt design to embed decision logic directly into the Planner Agent, enabling consistent interpretation, reasoning, and confirmation before execution.

The protocol comprises three functional components: 
\textit{Task Complexity Assessment}, \textit{Parameter Sufficiency Validation}, and \textit{Clarify-before-Execute}. Together, these components ensure that execution is always gated by input sufficiency, while reasoning depth adapts to task complexity. \textit{Task Complexity Assessment} determines how deeply the agent should reason about a user request. Each query is classified as \textit{Simple}, \textit{Complex}, or \textit{Agentic}. Simple tasks involve single-step execution with clear objectives (e.g., ``generate a network'' or ``run a simulation'') and are processed through direct responses without intermediate reasoning, while still verifying that all required parameters are provided. Complex tasks require multi-step analysis or comparison (e.g., ``analyze congestion'' or ``compare two traffic control policies'') and activate a \textit{Chain-of-Thought} reasoning template within \texttt{<thinking>} tags to guide systematic problem decomposition. Agentic tasks address strategic problems with multiple constraints and open-ended goals (e.g., ``find an optimal mitigation policy under a road closure'') and employ a multi-phase extended \textit{Chain-of-Thought} template, also using \texttt{<thinking>} tags, that expands reasoning to include solution-space exploration, trade-off analysis, validation criteria design, implementation planning, and synthesis of recommendations. This adaptive mechanism scales reasoning depth to task complexity, balancing computational efficiency with interpretability. \textit{Parameter Sufficiency Validation} ensures that the agent executes simulations only 
when essential information is available. Each request is mapped to a scenario schema covering spatial scope, demand or OD data, 
policy interventions, experimental design, and evaluation metrics. 
If any critical parameter is missing or ambiguous, the agent pauses execution and issues structured clarification questions with recommended defaults and explicit assumptions. Execution resumes only after all required parameters are confirmed, ensuring that each run is safe, reproducible, and well specified. \textit{Clarify-before-Execute} governs the interaction process when complex reasoning or missing parameters are detected. Before running a simulation, the agent summarizes the intended plan—inputs, steps, expected outputs, and validation checks—and waits for explicit user confirmation. 
Once approved, the plan is executed using the verified configuration, while simple tasks with sufficient parameters proceed directly with documented assumptions. This interactive process transforms the workflow from a command-driven sequence into a collaborative, goal-oriented scenario design loop, ensuring transparency, user control, and consistent behavior across sessions. Overall, these components enhance accessibility for non-expert users, increase transparency through explicit plan previews and confirmations, and improve computational efficiency by matching reasoning depth to task complexity.

\subsubsection{Supporting Infrastructure for Interactive Simulation}
\label{sec:infrastructure}

The Planner Agent operates on three supporting infrastructure components 
that enable efficient multi-turn interaction and stable deployment of the simulation workflow.

\begin{itemize}
    \item \textbf{Multi-turn Conversation Management.} 
          Interactive planning typically involves multiple rounds of parameter elicitation and refinement between the agent and the user. AgentSUMO maintains conversation history following multi-turn structure, where tool invocations and their results are embedded within the message sequence to preserve execution context. This design allows users to issue incremental commands such as ``now optimize traffic signals'' without re-specifying prior configurations, as the agent continuously retains awareness of the current network, routes, and simulation state throughout the session.
    
    \item \textbf{Simulation State Management.} 
          To support iterative refinement, AgentSUMO maintains a state dictionary that records simulation artifacts generated during the session, including network files, route configurations, and output results. Before executing new requests, the agent checks whether relevant networks, routes, or previous results already exist, preventing redundant computation and unnecessary re-generation. Each state update is automatically reflected in an XML context injected into the next system prompt, allowing the agent to make decisions in a \textit{state-aware} manner. This mechanism enables cumulative scenario development, where users progressively apply multiple interventions—such as lane adjustments or policy changes—without manual file handling.
    
    \item \textbf{Prompt Caching Optimization.} 
          The system prompt encodes domain knowledge, reasoning protocols, and tool interfaces required for SUMO operations. 
          To reduce computational overhead and improve responsiveness in extended interactions, a caching mechanism is applied to static sections of the prompt and tool definitions. Prompt caching is an increasingly common optimization technique in LLM systems that reuses preprocessed or semantically equivalent prompt segments across dialogue turns to avoid redundant token processing \citep{gim2024prompt}. This approach minimizes redundant processing of repeated context across turns, thereby improving runtime efficiency without altering reasoning behavior. Such optimization is critical for production-scale deployment, where multi-turn dialogue sessions generate a large number of iterative planning and analysis cycles.
\end{itemize}

These infrastructure components ensure that interactive simulations remain responsive, reproducible, and scalable across multi-turn sessions, providing a stable foundation for the Planner Agent’s reasoning and execution processes.

\subsection{Tool Layer via Model Context Protocol}

\subsubsection{Model Context Protocol Architecture}

The Model Context Protocol (MCP) addresses fragmentation in LLM–tool integration by providing a standardized, protocol-level abstraction for communication between AI applications and external resources~\citep{anthropic2024mcp}. Recently proposed by Anthropic and adopted across emerging development platforms such as Cursor, MCP formalizes model-to-tool interoperability through a unified transport and schema negotiation layer. Traditional integration approaches encounter significant limitations: manual API wiring necessitates custom authentication and data transformation for each service, platform-specific plugins such as ChatGPT Plugins impose ecosystem lock-in, and agent frameworks including LangChain enforce framework-dependent interfaces that restrict cross-platform compatibility. Drawing inspiration from the Language Server Protocol (LSP), MCP introduces several architectural innovations: (i)~protocol-based decoupling of tool implementation from usage, enabling dynamic publication of external functions independent of specific models or frameworks, (ii)~dynamic discovery and schema negotiation allowing runtime tool enumeration without prior configuration, (iii)~bi-directional communication channels supporting both model-to-tool requests and tool-initiated events, and (iv)~first-class access control and capability negotiation as protocol primitives, establishing foundations for auditable and secure AI-to-tool interactions ~\citep{hou2025modelcontextprotocolmcp}.

MCP employs a three-tier architecture comprising host, client, and server components. The \textbf{host} represents the AI application environment that executes agent-based tasks, such as Claude Desktop or Cursor, providing the runtime context for model interaction. The \textbf{client} functions as an intermediary within the host, maintaining a one-to-one connection with each MCP server while managing tool discovery, schema retrieval, and response processing. The \textbf{server} exposes external tools and resources through standardized capability declarations, processing invocation requests and returning structured results. Communication between client and server operates through a transport layer supporting local (STDIO) and remote (Server-Sent Events) protocols. The operational workflow follows a structured request-response pattern: the client queries the server to enumerate available tools, selects appropriate capabilities according to task context, executes tool invocations with validated parameters, and integrates returned results into the agent's reasoning process.

AgentSUMO implements this architecture to establish clear separation between AI reasoning logic and domain-specific SUMO utilities. The host layer employs the Claude API (Sonnet 4.5) as the language model backend, furnishing natural language understanding and multi-step planning capabilities. The client component manages the MCP protocol lifecycle through server discovery, tool schema conversion, and session coordination. The server provides a comprehensive suite of SUMO tools covering the complete simulation workflow from network preparation and scenario configuration to execution, analysis, and visualization. This architectural separation enables independent evolution of simulation capabilities and agent intelligence. Domain experts can extend SUMO functionality—such as incorporating new traffic signal optimization algorithms or vehicle emission models—without modifying agent code. Conversely, enhancements to the agent's planning logic, such as improved tool selection or multi-step reasoning strategies, proceed independently of simulation server implementation. The modular server design facilitates extensibility, as new traffic analysis capabilities can be registered without altering client code or model configuration. Furthermore, the standardized MCP interface supports multi-server orchestration, allowing AgentSUMO to integrate complementary capabilities such as filesystem MCP alongside SUMO tools, thereby enabling more sophisticated workflow composition for complex traffic policy experiments. To further support systematic analysis, AgentSUMO also incorporates a SQLite MCP server that stores SUMO simulation output files in a structured database. This integration enables efficient result querying and cross-scenario comparison, strengthening the framework’s capacity for large-scale post-simulation analysis and reproducible experiment management.

\subsubsection{Tool Integration}
                                              
MCP provides a lightweight JSON-RPC interface that standardizes communication between LLMs and external tools. By publishing machine-readable schemas, tools expose their capabilities in a format that allows LLMs to automatically discover functions and ensure validity before execution. AgentSUMO implements this protocol using FastMCP, wrapping SUMO functionalities as modular MCP tools. Each function is defined as a stateless Python tool registered with the MCP server via the \texttt{@mcp.tool()} decorator, which translates its signature and docstring into a structured schema. Beyond specifying parameters and return types, each tool's natural language description acts as a guidance to the LLM in reasoning about when and how the tool should be invoked. At runtime, the agent selects appropriate tools and instructs the client to execute them with schema-validated parameters through the MCP server.

\subsection{Agentic Simulation Workflow}
\label{sec:workflow}

Building on the Interactive Planning Protocol and the MCP tool layer, AgentSUMO operationalizes these concepts through an integrated workflow that unifies reasoning, tool invocation, and data management within SUMO. The workflow consists of three components—\textit{scenario generation}, \textit{policy experimentation}, and \textit{result analysis}—which together form an adaptive loop from user intent to interpretable insights.

\paragraph{Workflow Overview.}
AgentSUMO structures simulation tasks into goal-oriented workflows rather than rigid command sequences. When a user requests, for example, ``compare traffic before and after a road closure,'' the Planner Agent autonomously orchestrates network retrieval, baseline execution, policy intervention, comparative runs, and result synthesis without explicit step-by-step input. Adaptive reasoning adjusts computational depth to task complexity, selecting between direct tool calls and multi-step planning as appropriate. Parameter validation ensures completeness before execution, allowing non-expert users to refine goals interactively through clarification prompts. Supporting infrastructure—multi-turn conversation management, state-aware simulation tracking, and prompt caching—maintains execution context and improves efficiency across iterative sessions. Together these mechanisms enable collaborative, reproducible, and computationally efficient simulations.

\paragraph{Scenario Generation Component.}
Scenario generation establishes the foundational simulation environment by constructing a road network and generating corresponding trip demand. 
Network generation begins by defining the spatial extent of the simulation. When actual OD data are available, the agent extracts a bounding box from the given OD coordinates to define the target region. If no OD data are provided, the user can simply specify a location (e.g., ``Times Square, Manhattan'') and a radius, and the agent uses GeoPy to geocode the location, compute the corresponding bounding box. Once the spatial extent is defined, the agent retrieves OSM data for the region via \texttt{osmGet.py} and converts it into a SUMO network using \texttt{netconvert}, producing lane geometries, intersections, and signal definitions suitable for simulation.

Trip demand generation supports both random and actual OD modes. When the user does not have actual OD data, the agent generates random OD demand using \texttt{randomTrips.py} according to user-specified traffic conditions (e.g., light, medium, or heavy). Traffic volume is automatically scaled to the network size so that demand intensity remains proportional to the spatial extent. When the user provides actual OD data, the agent preprocesses it accordingly: coordinate-based inputs are map-matched to the network and converted into trip files, while zone-based OD matrices require a shapefile and are processed through \texttt{polyconvert}, \texttt{edgesInDistricts.py}, and \texttt{od2trips}. 

Routing is performed with \texttt{duarouter}, generating route files that are combined with the network to form a SUMO configuration that can be executed directly within the simulation environment. This integrated process abstracts SUMO’s complex data preparation into a single natural-language request, allowing non-experts to generate realistic and reproducible simulation scenarios without manual preprocessing.

\paragraph{Policy Experimentation Component.}
The policy experimentation phase enables evaluation of interventions in signal control, infrastructure, and traffic management. Signal control optimization includes \textit{Green Wave coordination}, which synchronizes signals along a corridor using \texttt{tlsCoordinator.py}, and \textit{Webster’s cycle adaptation}, which calculates optimal cycle lengths and phase splits via \texttt{tlsCycleAdaptation.py} \citep{cce62891b31b4e9ebf12040faa120deb}. Infrastructure tools modify network capacity through road closure, lane reduction, or speed adjustment, 
each implemented by XML editing or \texttt{netconvert} operations. 
Traffic management tools handle vehicle generation and fleet composition adjustments, allowing users to model demand perturbations or technology adoption scenarios. These seven modular tools can be combined within one conversational session, supporting comprehensive policy analysis without manual XML editing.

\paragraph{Result Analysis and Visualization Component.}
Result analysis converts simulation outputs into interpretable metrics and visual representations. The agent runs SUMO simulations to produce \texttt{tripinfo.xml}, \texttt{edgeData.xml}, and \texttt{summary.xml}. Parsed outputs are summarized using \texttt{attributeStats.py} and transformed into JSON with semantic labels 
(e.g., average travel time, congestion delay) for direct use in natural-language reports or Q\&A. Visualization utilities (\texttt{plot\_net\_selection.py}, \texttt{plot\_net\_dump.py}) generate 
spatial and temporal maps highlighting traffic density, speed, or emissions. Simulation results are also stored in a structured database through an SQLite MCP server, enabling efficient querying and cross-scenario comparison for large-scale analysis.

Across all components, AgentSUMO demonstrates how reasoning protocols, standardized tool integration, and state-aware management converge into a unified agentic framework for interactive traffic simulation. 
The workflow lowers the barrier for scenario design, enhances reproducibility, and supports scalable analysis from high-level policy intent to actionable insight.

\section{Experimental Results}\label{sec:results}

\subsection{Setup}\label{sec:setup}
All experiments were performed on Ubuntu 20.04.1 LTS running on an Intel Core i9-10900X CPU with two NVIDIA RTX 3090 GPUs (24GB each) and 128GB RAM. We used Python 3.12 with MCP 1.13.1 and Eclipse SUMO 1.24.0. The AgentSUMO was powered by Anthropic's \texttt{Claude Sonnet 4.5} model, executed via the \texttt{FascMCP} runtime. 

Two datasets were employed in this study. First, publicly available \textit{Yellow Cab} trip records from the \textbf{New York City Taxi and Limousine Commission}\footnote{\url{https://www.nyc.gov/site/tlc/about/tlc-trip-record-data.page}} were used as empirical data. The dataset comprises 32,480 trips within Manhattan, recorded on Monday, April~4,~2016, between 19:00 and 20:00. Each record includes pick-up and drop-off coordinate locations and times.
Second, anonymized taxi trajectory data were obtained under a non-disclosure agreement from the \textbf{Korea~Transport~Institute~(KOTI)}. This dataset consists of 10 second interval GPS logs from 68,310 registered taxis in Seoul during April~2018. For privacy protection, the raw trajectory data were aggregated and transformed into coordinate-based OD pairs rather than full trajectories. In this study, trips recorded between 07:00 and 09:00 on Monday, April~2,~2018 within the Gangnam district of Seoul, comprising 6,200 trips, were used. Due to confidentiality agreements, this dataset cannot be publicly released.
 
 Performance was evaluated using metrics from SUMO simulation outputs. Trip-level metrics from \texttt{tripinfo.xml} include \textit{duration} (s), the time each vehicle needed to accomplish its trip; \textit{time loss} (s), the time lost due to driving below ideal speed; \textit{CO$_2$\_abs} (mg), \textit{PMx\_abs} (mg), and \textit{NOx\_abs} (mg), the complete amounts of carbon dioxide, particulate matter, and nitrogen oxides emitted per trip; \textit{fuel\_abs} (mg) and \textit{electricity\_abs} (Wh), measuring total fuel and electricity consumed per trip. Edge-level metrics from \texttt{edgeData.xml} include \textit{density} (veh/km), the vehicle density on the edge; \textit{occupancy} (\%), the percentage of edge space occupied by vehicles; and \textit{speed} (m/s), the space-mean speed averaged over time and space. These metrics were aggregated via SUMO's \texttt{attributeStats.py} through averaging or summing as appropriate. All simulations were executed under identical configurations to ensure comparability across scenarios.


\subsection{Experiment 1: Simple Tasks}
Simple tasks in AgentSUMO represent low-complexity simulation goals that can be achieved through single-step reasoning and direct tool invocation. 
These tasks typically involve straightforward scenario generation such as: \textit{"Run a traffic simulation around Gangnam Station in Seoul within a 2~km radius."} 
Upon receiving such a query, the agent performs a parameter sufficiency check to verify that all essential inputs are specified. 
If key parameters—such as trip type, traffic condition, or simulation duration—are missing, AgentSUMO initiates a brief clarification dialogue, proposes default values, and confirms user intent. 
Once validated, the system automatically generates the SUMO network, routes, and configuration files through its MCP-based orchestration pipeline. Using this workflow, AgentSUMO was tested in two urban contexts: Gangnam Station in Seoul and Times Square in New York. 
For each site, a 2~km-radius subnetwork was extracted from OSM, and simulations were executed under three traffic conditions—\textit{light}, \textit{medium}, and \textit{heavy}—with trip demand generated by the RandomOD module.

\begin{figure}[pos=t]
    \centering
    \begin{subfigure}[pos=t]{0.95\textwidth}
        \centering
        \includegraphics[width=\textwidth]{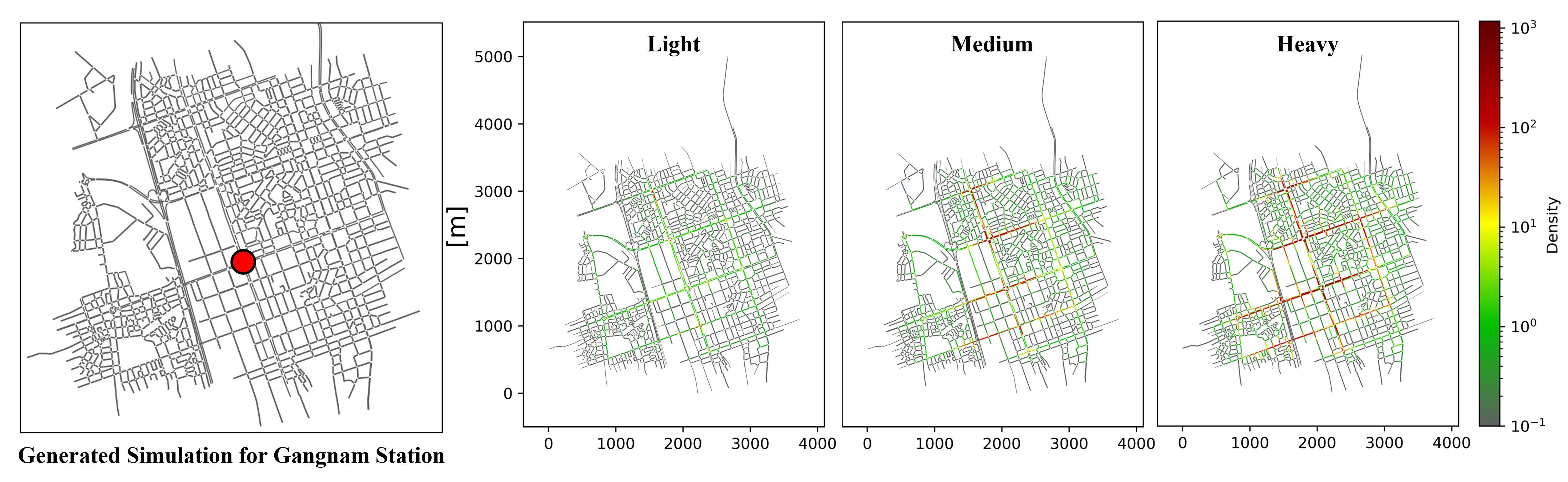}
        \caption{Generated simulation and traffic density heatmap for \textbf{Gangnam Station}.}
        \label{fig:exp1_gangnam}
    \end{subfigure}

    \vspace{1em}

    \begin{subfigure}[pos=t]{0.95\textwidth}
        \centering
        \includegraphics[width=\textwidth]{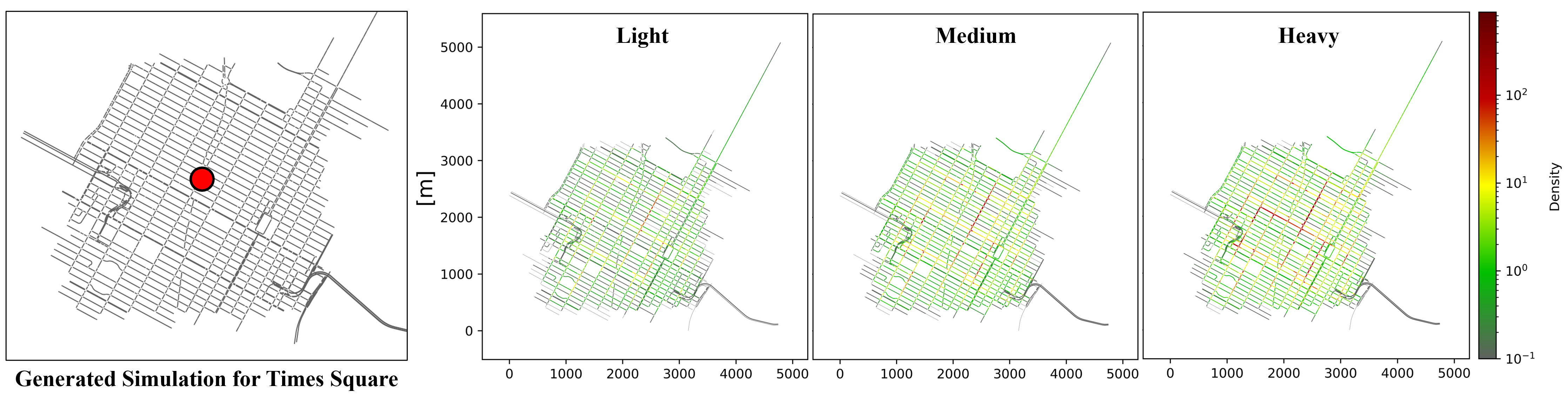}
        \caption{Generated simulation and traffic density heatmap for \textbf{Times Square}.}
        \label{fig:exp1_timesquare}
    \end{subfigure}

    \caption{Qualitative results of Simple Tasks across two regions under varying traffic conditions.}
    \label{fig:exp1_results}
\end{figure}

The generated networks accurately captured local topology and showed consistent variations in traffic density across light, medium, and heavy demand conditions (Figure~\ref{fig:exp1_results}). 
Table~\ref{tab:exp1_results} summarizes the key performance indicators—average density, travel time, time loss, and CO$_2$ emissions—under each condition. 
Metric values increased proportionally with demand intensity, confirming that the simulation behaved as expected under scaled traffic loads. 
The experiment demonstrates that AgentSUMO can autonomously construct valid SUMO configurations, verify parameter sufficiency, and execute reproducible simulations with minimal user input. 
This establishes a baseline for subsequent experiments involving higher-level policy evaluations and multi-step reasoning tasks.

\begin{table*}[htbp]
\centering
\caption{Quantitative results of Simple Tasks across different traffic conditions.}
\label{tab:exp1_results}
\renewcommand{\arraystretch}{1.4}
\begin{tabular*}{\tblwidth}{@{\extracolsep{\fill}}llcccc@{}}
\toprule
\textbf{Region} & \textbf{Condition} & \textbf{Density (veh/km)} & \textbf{Duration (s)} & \textbf{Time Loss (s)} & \textbf{CO2 Emission (kg)} \\
\midrule
\multirow{3}{*}{Gangnam Station} 
  & Light (1587 veh/h)   & 2.42  & 256.38 & 153.15 & 1.01 \\
  & Medium (4230 veh/h)  & 6.63 & 418.31 & 318.30 & 1.38 \\
  & Heavy (6874 veh/h)   & 12.06 & 415.26 & 518.58 & 1.57 \\
\addlinespace
\midrule
\multirow{3}{*}{Times Square} 
  & Light (1403 veh/h)   & 11.64  & 430.26 & 293.27 & 1.38 \\
  & Medium (3234 veh/h)  & 25.98 & 440.79 & 303.21 & 1.43 \\
  & Heavy (6078 veh/h)   & 39.05 & 489.38 & 352.44 & 1.50 \\
\bottomrule
\end{tabular*}
\end{table*}


\subsection{Experiment 2: Complex Tasks}

\begin{figure}[pos=t]
  \centering
  \includegraphics[width=\linewidth,height=0.8\textheight,keepaspectratio]{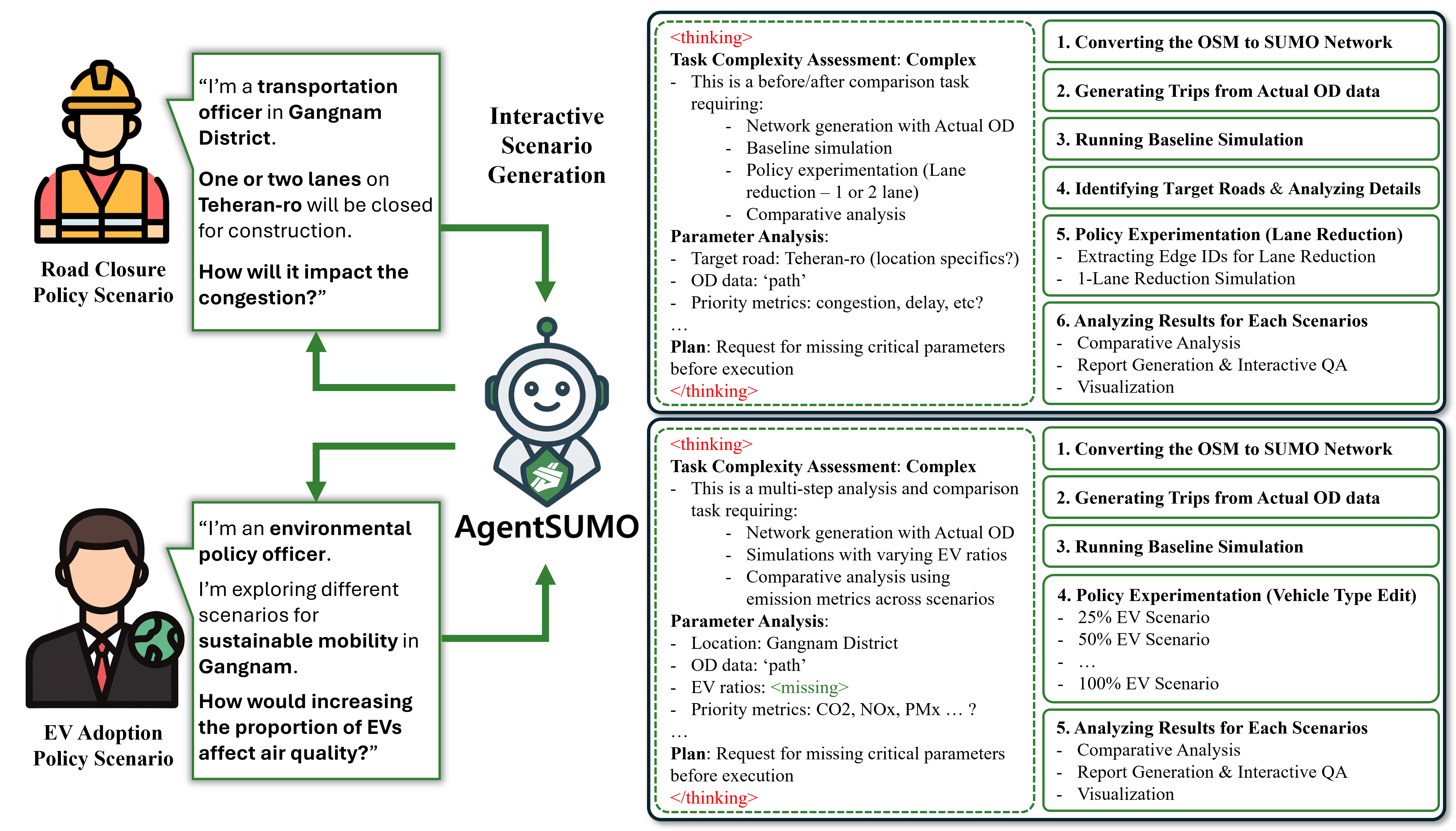}
  \caption{llustration of complex task reasoning and execution pipeline in AgentSUMO (lane-closure and EV-adoption scenarios)} 
  \label{fig:exp2-overview}
\end{figure}

Complex tasks in AgentSUMO require multi-step reasoning and contextual understanding to perform comparative or evaluative analyses rather than single-run simulations. 
As summarized in Figure~\ref{fig:exp2-overview}, the agent decomposes user intent into sequential phases of data validation, scenario setup, execution planning, and result interpretation before invoking SUMO tools.

Two representative policy experiments were conducted in the Gangnam District to evaluate AgentSUMO’s multi-step reasoning capability. 

\textbf{(1) Road-closure scenario.} 
A transportation officer requested: \textit{“One or two lanes on Teheran-ro will be closed for construction. How will this affect congestion?”} 
Since the query lacked details on the location and number of lanes, the agent initiated a \textit{clarify-before-execute} dialogue to confirm the affected segment, number of closed lanes, OD data type, and simulation time window (07:00–09:00). 
After verification, AgentSUMO constructed paired simulation scenarios—\textit{pre-construction} and \textit{post-construction}—and executed both under identical conditions. 
The resulting outputs included aggregated performance metrics and congestion heatmaps for the Teheran-ro corridor.

\begin{figure}[pos=t]
  \centering
  \includegraphics[width=0.85\linewidth,height=0.85\textheight,keepaspectratio]{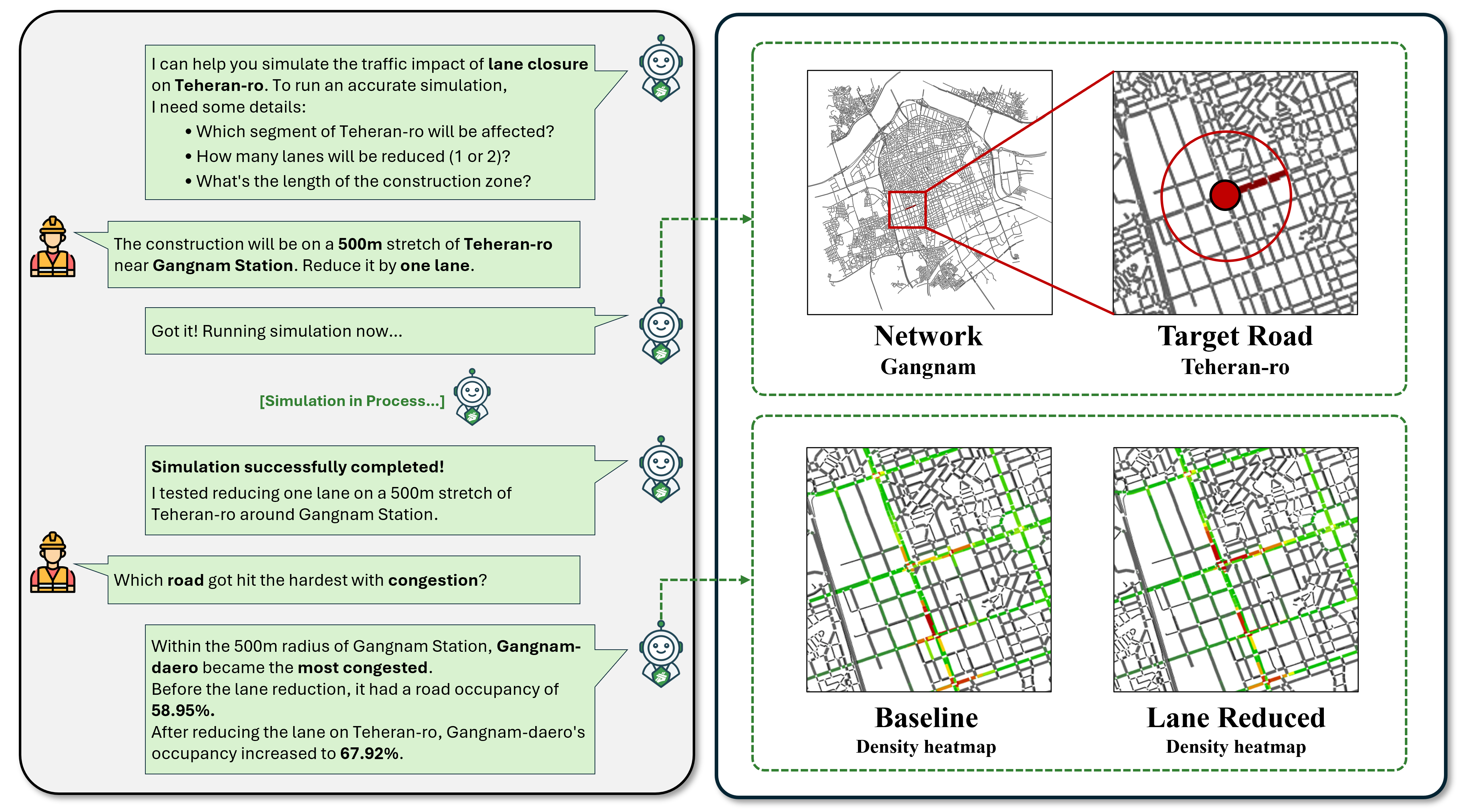}
  \caption{Interactive scenario generation and executed outcome for the lane-closure scenario on Teheran-ro.}
  \label{fig:exp2_result1}
\end{figure}

Figure~\ref{fig:exp2_result1} and Table~\ref{tab:exp2_result1} summarize this workflow and outcome. 
The left panel of Figure \ref{fig:exp2_result1} illustrates how the agent identified missing parameters, requested clarification, and generated interpretable feedback. 
Following user confirmation of a 500~m construction zone near Gangnam Station with one lane closed, the simulation results showed higher congestion around Gangnam-daero, where average road occupancy increased from 58.95\% to 67.92\%. 
Network-wide density rose from 7.48 to 7.68~veh/km, mean travel duration from 441.66 to 485.00~s, and time loss from 325.79 to 368.76~s, while average speed slightly decreased from 9.19 to 9.08~m/s. 
These findings indicate measurable performance degradation due to lane reduction and demonstrate the system’s ability to refine underspecified user intents and execute controlled comparative simulations.

\begin{figure}[pos=t]
  \centering
  \begin{subfigure}{0.85\linewidth}
    \centering
    \includegraphics[width=\linewidth,height=0.85\textheight,keepaspectratio]{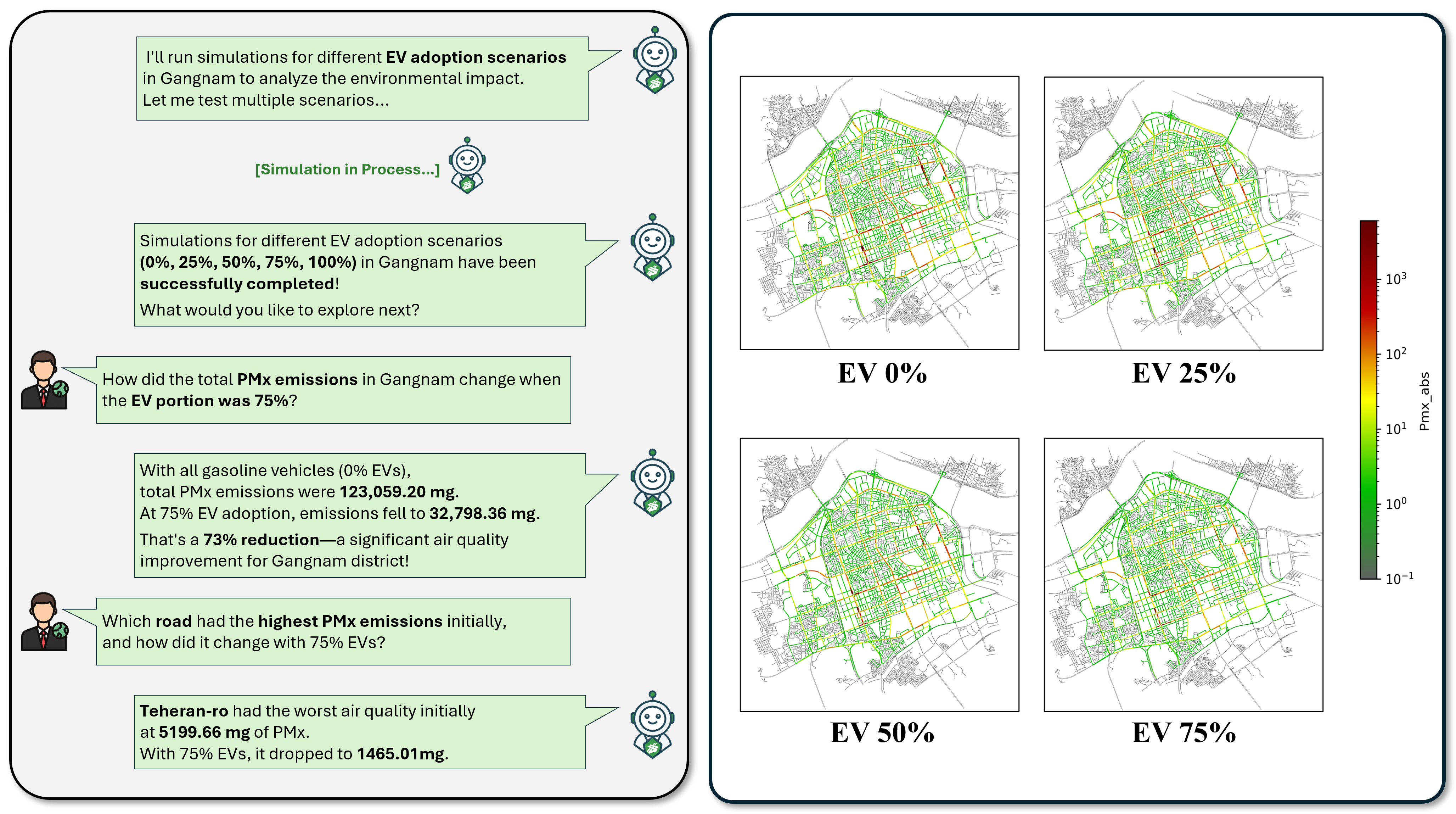}
    \caption{Interactive scenario generation and comparison of average PM$_x$ across EV ratios on the Gangnam network.}
    \label{fig:exp2_result2_1}
  \end{subfigure}

  \vspace{1em}
  
  \begin{subfigure}{0.8\linewidth}
  \centering
    \includegraphics[width=\linewidth,height=0.85\textheight,keepaspectratio]{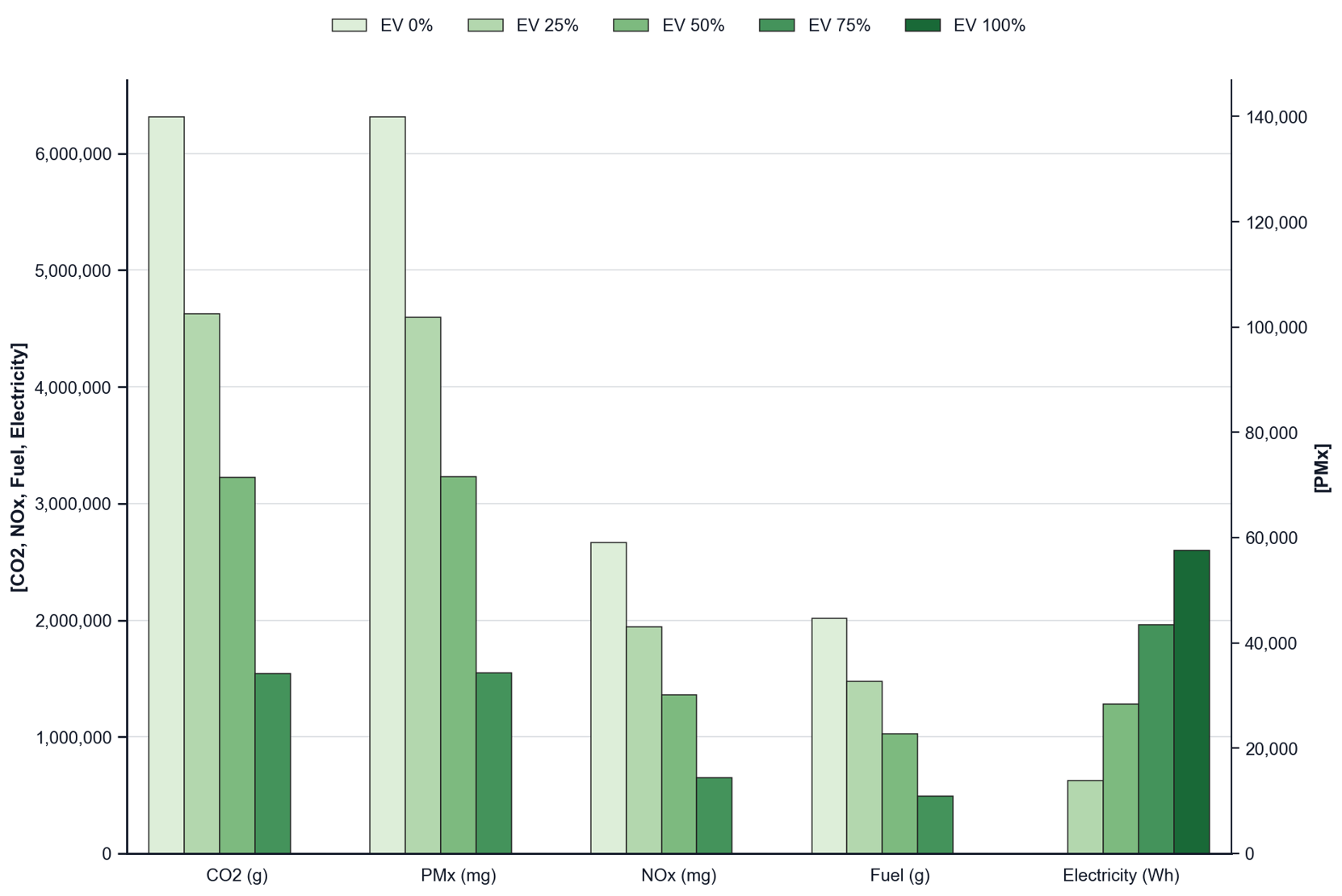}
    \caption{Comparative emission metrics (CO$_2$, PM$_x$, NO$_x$, Fuel, and Electricity) across EV ratios (0–100\%).}
    \label{fig:exp2_result2_2}
  \end{subfigure}

  \caption{Results of the EV adoption policy scenario in AgentSUMO.}
  \label{fig:exp2_result2}
\end{figure}

\textbf{(2) EV-adoption scenario.} 
An environmental policy officer asked: \textit{“How would increasing the proportion of electric vehicles affect air quality in Gangnam?”} 
AgentSUMO verified parameter completeness, identified missing values such as EV proportion and emission metrics, and proposed default increments (0\%, 25\%, 50\%, 75\%, 100\%) with indicators (CO$_2$, NO$_x$, PM$_x$) for confirmation. 
The left panel of Figure~\ref{fig:exp2_result2_1} illustrates the agent’s interactive reasoning, while the right panel (Figure~\ref{fig:exp2_result2_1}) visualizes spatial emission reductions across the network. 
Simulation results showed that total PM$_x$ emissions in Gangnam decreased from 123{,}059.20~mg at the 0\% baseline to 32{,}798.36~mg at 75\% EV adoption—a 73\% reduction. 
The largest improvement occurred along Teheran-ro, where PM$_x$ levels declined from 5{,}199.66~mg to 1{,}465.01~mg. 
Figure~\ref{fig:exp2_result2_2} presents aggregated metrics, confirming consistent decreases in CO$_2$, NO$_x$, PM$_x$, and fuel consumption, alongside a proportional rise in electricity usage with higher EV penetration.

Collectively, these results demonstrate that AgentSUMO can autonomously construct, execute, and interpret multi-scenario simulations through natural-language interaction, providing transparent and reproducible analyses for non-expert users.

\begin{table}[htbp]
\centering
\caption{Quantitative results for the lane-closure scenario}
\renewcommand{\arraystretch}{1.4}
\setlength{\tabcolsep}{18pt}      
\begin{tabular}{l c c c c}
\toprule
\textbf{Type} & \textbf{Density (veh/km)} & \textbf{Duration (s)} & \textbf{Time Loss (s)} & \textbf{Speed (m/s)} \\
\midrule
Pre-Construction      & 7.48  & 441.66 & 325.79 & 9.19 \\
Post-Construction   & 7.68 & 485.00 & 368.76 & 9.08 \\
\bottomrule
\end{tabular}
\label{tab:exp2_result1}
\end{table}

\subsection{Experiment 3: Agentic Tasks}

\begin{figure}[pos=t]
  \centering
  \includegraphics[width=\linewidth,height=0.8\textheight,keepaspectratio]{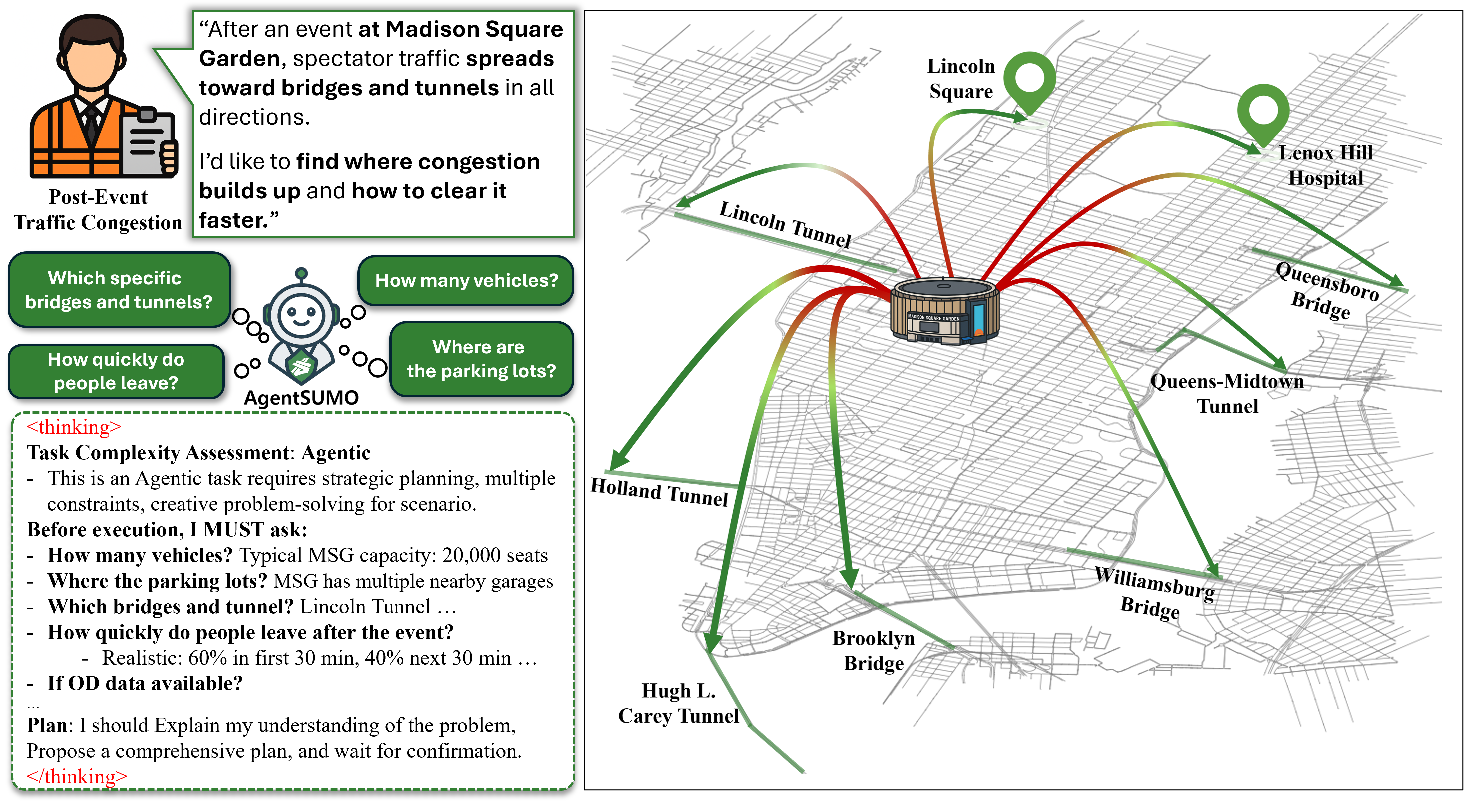}
  \caption{Illustration of agentic task reasoning and interactive planning in AgentSUMO (post-event congestion scenario at Madison Square Garden)} 
  \label{fig:exp3_overview}
\end{figure}

Agentic tasks represent the highest level of reasoning complexity in AgentSUMO, requiring strategic planning, multi-constraint decision-making, and creative problem solving under uncertainty. 
Unlike simple or complex tasks, these tasks require the system to formulate the problem, generate feasible simulation hypotheses, and iteratively refine its execution plan through user clarification and contextual feedback.

A transportation management officer requested: 
\textit{“After an event at Madison Square Garden, spectator traffic spreads toward bridges and tunnels in all directions. 
I’d like to find where congestion builds up and how to clear it faster.”} 
Since the query involved multiple uncertainties—such as the number of departing vehicles, parking locations, and egress rate—AgentSUMO initiated a \textit{clarify-before-execute} reasoning loop to identify missing parameters, propose reasonable assumptions, and validate them through user dialogue before generating the scenario. 
This process reflects the agent’s ability to decompose an ill-defined problem into structured simulation tasks, forming the basis of an agentic reasoning workflow.

This experiment focuses on post-event traffic management near Madison Square Garden (MSG) in Manhattan, where spectator traffic disperses toward major bridges and tunnels in multiple directions. 
In such cases, the agent must integrate contextual factors---event capacity, parking availability, egress rate, and OD distribution---to design a realistic evacuation plan. 
Through its interactive reasoning loop, AgentSUMO identifies missing information, proposes assumptions, validates them through dialogue, and constructs a comprehensive multi-phase simulation workflow.

Nine representative destination zones were defined (Figure~\ref{fig:exp3_overview}), corresponding to major bridges, tunnels, and residential districts across Manhattan. 
These include the Lincoln Tunnel, Holland Tunnel, Queensboro Bridge, and Williamsburg Bridge, capturing outbound flows toward New Jersey, Queens, and Brooklyn, reflecting realistic post-event evacuation patterns observed in central Manhattan.

To construct the OD configuration, three origin points were selected near MSG: Madison Square Garden, PENN~1, and Penn South Playground (PSP). 
Assuming an event capacity of 20{,}000 spectators, approximately 2{,}000 vehicles (10\%) were generated. 
Each origin was connected to multiple destinations as summarized in Table~\ref{tab:exp3_od}. 
For each OD pair, 60\% of vehicles departed within the first 30~minutes after the event, and the remaining 40\% departed gradually until the end of the simulation period, representing a realistic egress profile.

\begin{figure}[pos=t]
  \centering
  \includegraphics[width=\linewidth,height=0.8\textheight,keepaspectratio]{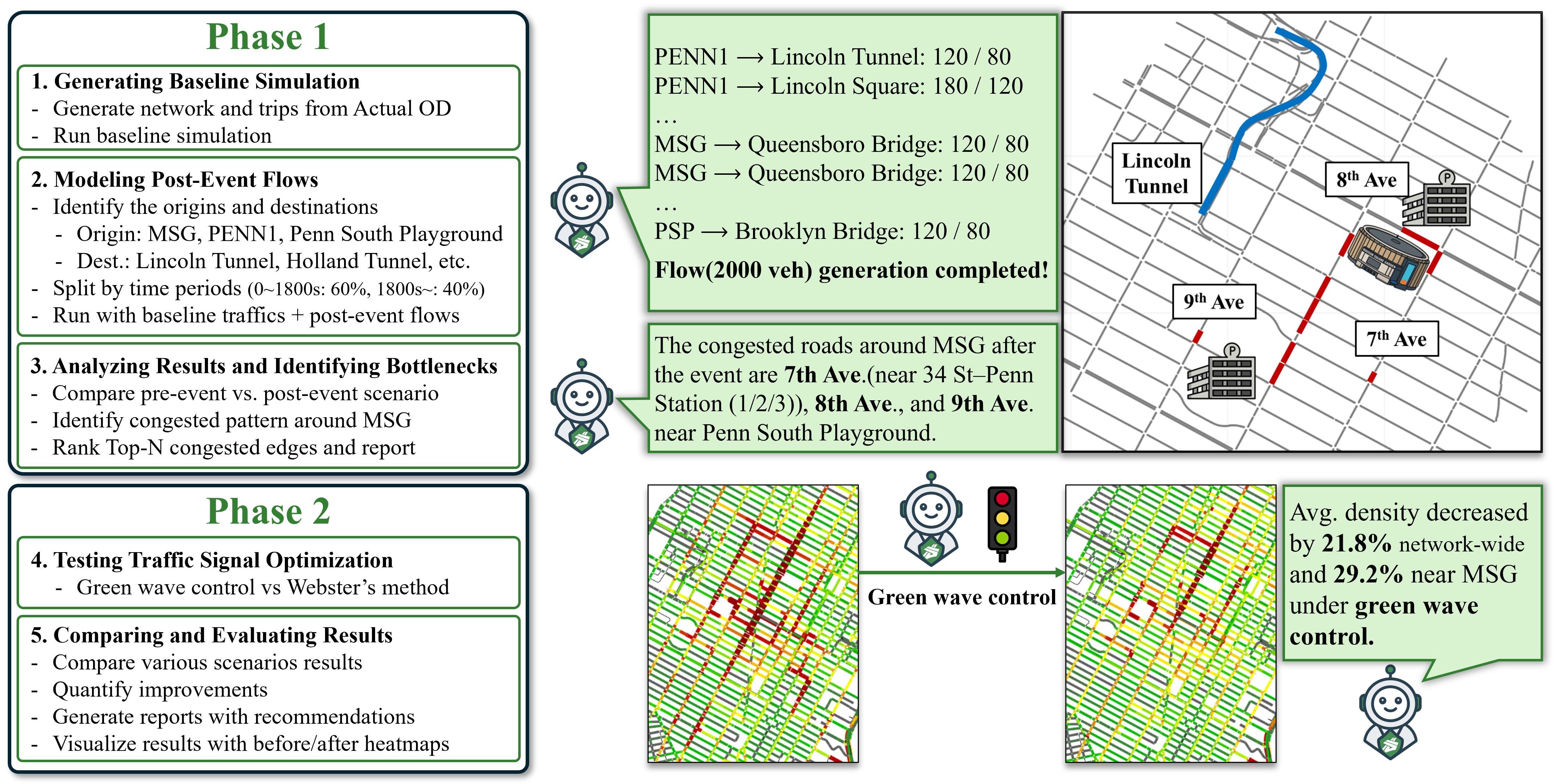}
  \caption{Task planning and executed outcomes for the MSG post-event scenario} 
  \label{fig:exp3_result1}
\end{figure}

\begin{table}[htbp]
\centering
\caption{Origin--destination design for the Madison Square Garden post-event traffic scenario. 
The numbers in parentheses indicate vehicle counts during the first and second half of the simulation.}
\renewcommand{\arraystretch}{1.2}
\begin{tabular}{lllcc}
\toprule
\textbf{Origin} & \textbf{Destination} & \textbf{Outbound Direction / Area} & \textbf{Vehicles} & \textbf{Initial / Later Period} \\
\midrule
\multirow{3}{*}{PENN~1} 
 & Lincoln Tunnel        & Westbound -- Toward New Jersey   & 200 & 120 / 80 \\
 & Lincoln Square        & Northbound -- Manhattan interior     & 300 & 180 / 120 \\
 & Lenox Hill Hospital   & Northeast -- Upper East Side, Manhattan & 300 & 180 / 120 \\
\midrule
\multirow{3}{*}{MSG} 
 & Williamsburg Bridge   & Southeast -- Toward Brooklyn        & 200 & 120 / 80 \\
 & Queens--Midtown Tunnel & Eastbound -- Toward Queens          & 200 & 120 / 80 \\
 & Queensboro Bridge     & Northeast -- Toward Queens          & 200 & 120 / 80 \\
\midrule
\multirow{3}{*}{PSP} 
 & Holland Tunnel        & Southwest -- Toward New Jersey      & 200 & 120 / 80 \\
 & Hugh~L.~Carey Tunnel  & Southbound -- Toward Brooklyn            & 200 & 120 / 80 \\
 & Brooklyn Bridge       & Southeast -- Toward Brooklyn        & 200 & 120 / 80 \\
\bottomrule
\end{tabular}
\label{tab:exp3_od}
\end{table}

As illustrated in Figure~\ref{fig:exp3_result1}, this experiment demonstrates AgentSUMO’s agentic reasoning and execution workflow for the MSG post-event scenario. 

\textbf{Phase 1: Baseline and post-event flow analysis.}  
As shown in the upper panel of Figure~\ref{fig:exp3_result1}, AgentSUMO first constructed the network and route files based on the Actual OD data and executed a baseline simulation to capture ordinary traffic conditions across Manhattan. 
It then incorporated post-event flows from Table~\ref{tab:exp3_od}, representing vehicles dispersing from the three origins toward nine major destinations across Manhattan, New Jersey, Queens, and Brooklyn. 
By combining the baseline traffic with event-related demand, AgentSUMO identified congestion hotspots around MSG and reported them through natural-language summaries, highlighting severe congestion along 7th, 8th, and 9th Avenues.

\textbf{Phase 2: Mitigation policy evaluation.}  
As illustrated in the lower panel of Figure~\ref{fig:exp3_result1}, the agent searched for optimal signal control strategies to alleviate the severe congestion identified in Phase 1.
Two methods were compared—Green Wave coordination and Webster’s adaptive control. Simulations indicated that the Green Wave strategy produced the greatest improvement, reducing average network-wide density by approximately 22\% and local congestion near MSG by about 29\%. 
The strategy also outperformed Webster’s control along major arterials, improving flow continuity and reducing local bottlenecks.

\begin{figure}[pos=t]
  \centering
  
  \begin{subfigure}{0.9\linewidth}
  \centering
    \includegraphics[width=\linewidth,height=0.8\textheight,keepaspectratio]{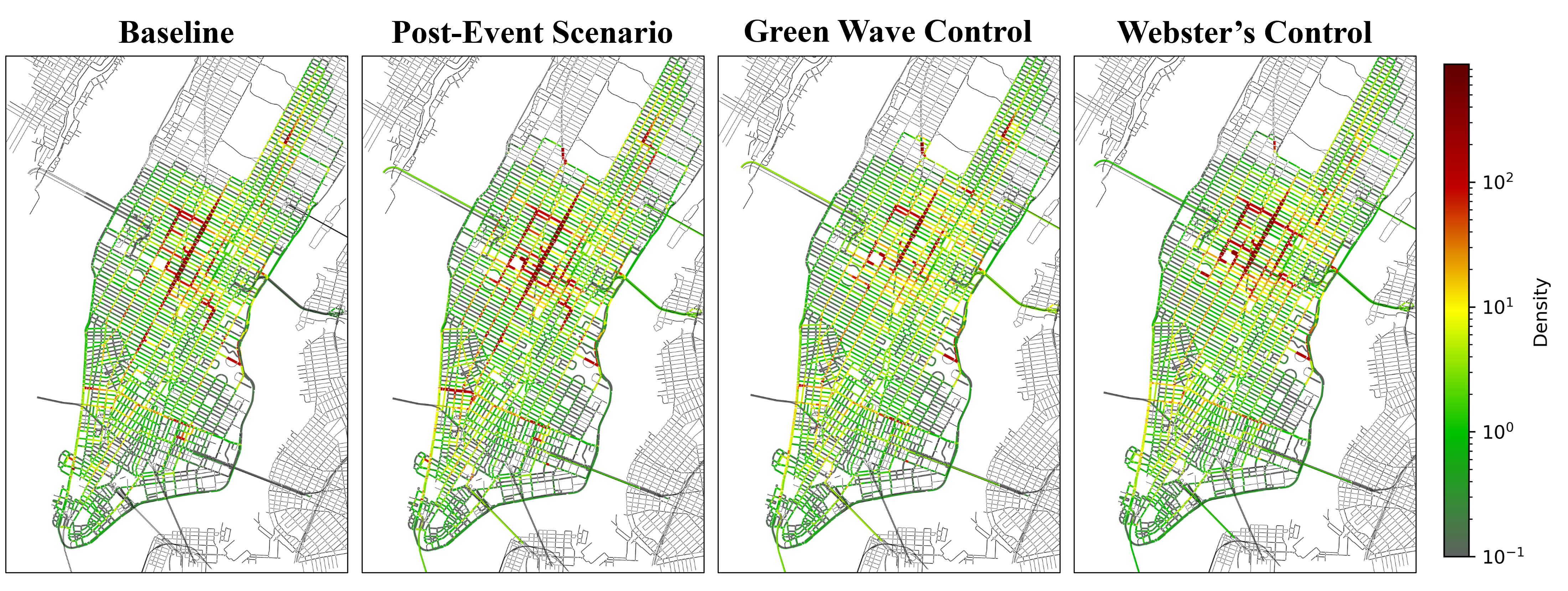}
    \caption{Comparison of average traffic density across scenarios on the Manhattan network.}
    \label{fig:exp3_result2_1}
  \end{subfigure}

  \vspace{1em}
  
  \begin{subfigure}{0.9\linewidth}
    \centering
    \includegraphics[width=\linewidth,height=0.8\textheight,keepaspectratio]{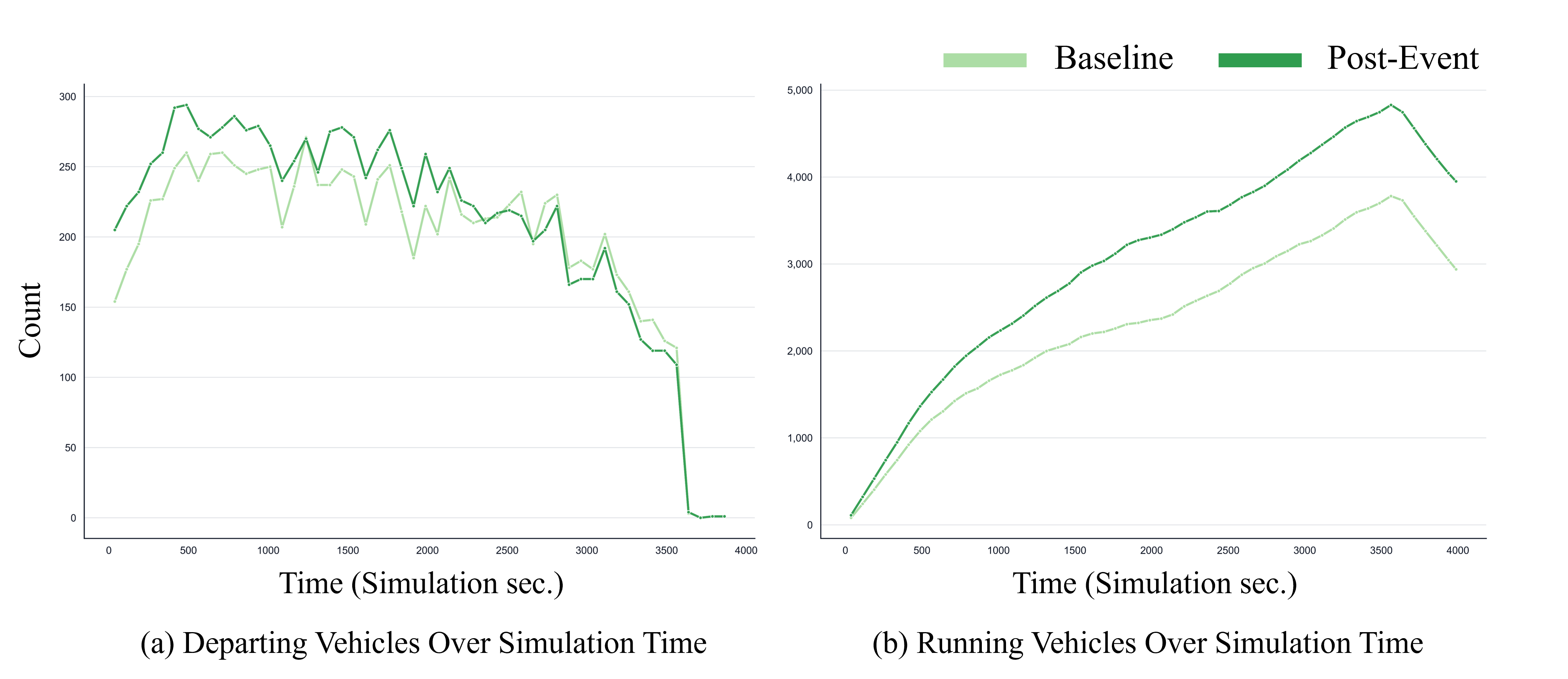}
    \caption{Comparison of pre-event and post-event traffic dynamics.}
    \label{fig:exp3_result2_2}
  \end{subfigure}

  \caption{Results of the MSG post-event scenario.}
  \label{fig:exp3_result2}
\end{figure}

\begin{table}[htbp]
\centering
\caption{Comparison of average traffic density under different signal control strategies in the post-event scenario.}
\renewcommand{\arraystretch}{1.6}
\setlength{\tabcolsep}{7pt}
\begin{tabular}{lcccc}
\toprule
\textbf{Scenario} & \textbf{Network-wide Avg.} & \textbf{Top 10 Congested Edges} & \textbf{Within 300\,m of MSG} & \textbf{Within 500\,m of MSG} \\
\midrule
Post-Event  & 16.5 & 501.0 & 73.3 & 59.0 \\
Green Wave  & \textbf{12.9} & \textbf{277.6} & \textbf{51.9} & \textbf{42.7}\\
Webster     & 15.2 & 373.1 & 69.3 & 57.3 \\
\bottomrule
\end{tabular}
\label{tab:exp3_density}
\end{table}

As illustrated in Figure~\ref{fig:exp3_result2_1}, incorporating post-event traffic significantly intensifies congestion in central Manhattan compared with ordinary conditions. 
Vehicle flows expand outward from Madison Square Garden toward multiple exit corridors, with particularly high densities observed near the Holland Tunnel and the northeastern outbound routes toward Queens. 
Among the two mitigation strategies tested, the Green Wave coordination effectively smooths traffic flow and alleviates localized bottlenecks, outperforming Webster’s adaptive control both in the event vicinity and along major arterials. 
These spatial patterns align with the quantitative results summarized in Table~\ref{tab:exp3_density}. 
Under Green Wave control, the network-wide average density decreased from 16.5 to 12.9\,veh/km, while congestion near Madison Square Garden dropped from 73.3 to 51.9\,veh/km within 300\,m and from 59.0 to 42.7\,veh/km within 500\,m. 
A similar trend was observed for the ten most congested edges, where average density declined by nearly 45\%, confirming that the Green Wave strategy substantially mitigated post-event congestion.

As shown in Figure~\ref{fig:exp3_result2_2}, the temporal dynamics further demonstrate that the post-event demand was effectively represented in the simulation. 
In the departing-vehicle profile (Figure~\ref{fig:exp3_result2_2}a), the sharp increase within the first 1{,}800~s corresponds to the modeled 60\% egress rate immediately after the event. 
Following this surge, the departure rate stabilizes—not due to reduced demand, but because network saturation near Madison Square Garden limits additional vehicle entry. 
The running-vehicle count (Figure~\ref{fig:exp3_result2_2}b) exhibits a similar trend: rapid accumulation, a temporary plateau as the network reaches capacity, and gradual dissipation as vehicles exit through major outbound routes. 
These results verify that AgentSUMO accurately captures both the temporal evolution and capacity-constrained behavior of post-event congestion.

\textbf{Summary.}  
Collectively, the findings in Figures~\ref{fig:exp3_result2_1}–\ref{fig:exp3_result2_2} and Table~\ref{tab:exp3_density} demonstrate AgentSUMO’s ability to autonomously reason through complex, ill-defined traffic management problems and synthesize effective mitigation strategies through iterative simulation and feedback. 
By combining contextual understanding, adaptive planning, and multi-scenario evaluation, AgentSUMO functions as an agentic framework for realistic, decision-supportive urban traffic simulation.

\section{Conclusion}\label{sec:conclusion}

In this paper, we proposed AgentSUMO, an agentic framework for interactive simulation scenario generation in SUMO via LLM. 
By positioning the LLM as a reasoning and orchestration agent, AgentSUMO enables non-expert users to design, execute, and analyze realistic traffic simulations through natural-language interaction. 
The framework integrates the Interactive Planning Protocol for adaptive reasoning and MCP for standardized tool orchestration, forming an end-to-end workflow that transforms abstract user goals into executable SUMO experiments. 
Through a series of experiments—spanning Simple, Complex, and Agentic task types—AgentSUMO demonstrated its capacity for autonomous scenario construction, comparative policy evaluation, and multi-phase reasoning under real-world conditions. 
Across these tasks, the framework consistently produced interpretable, reproducible, and policy-relevant outcomes, thereby bridging the gap between expert simulation workflows and accessible, data-driven decision support for urban mobility planning.

Building on these findings, several research directions emerge to further advance agentic simulation frameworks in urban mobility. 
Future work will extend AgentSUMO toward foundation-model generalization, exploring how diverse LLM backends can be integrated through the same MCP interface to enhance robustness and adaptability. 
Another avenue lies in scaling the framework to larger and multimodal networks, where coordinated reasoning across road, public-transit, and shared-mobility layers can be examined. 
To strengthen real-world applicability, we also plan to incorporate human-in-the-loop evaluation with planners and policymakers, assessing how interactive reasoning can augment decision quality and transparency in practice. 
Finally, we aim to evolve AgentSUMO as a broader foundation-model infrastructure for intelligent urban simulation, coupling traffic dynamics with environmental, safety, and energy-policy domains. 
Rather than addressing limitations, these directions represent the natural progression of AgentSUMO toward a scalable, extensible, and decision-supportive platform for next-generation urban mobility research.









\printcredits

\section*{Acknowledgements}
This work was supported by the National Research Foundation of Korea(NRF) grant funded by the Korea government(MSIT) (No. 2022M3J6A1063021, No. RS-2025-00517342) This research was also supported by Center for Advanced Urban Systems (CAUS) of Korea Advanced Institute of Science and Technology (KAIST) funded by GS E\&C.

\section*{Declaration of Competing Interest}
The authors declare that they have no known competing financial interests or personal relationships that could have appeared to influence the work reported in this paper.

\section*{Data Availability}
The data that support the findings of this study are partly publicly available and partly not publicly available under a non-disclosure agreement. 

\bibliographystyle{cas-model2-names}

\bibliography{AgentSUMO_refs}



\end{document}